\documentclass{article}
\usepackage{spconf,amsmath,graphicx,bm}
\usepackage{amsfonts} 
\usepackage{mathtools}
\usepackage{color}
\usepackage{cite}
\usepackage{amsthm}

\newtheorem{theorem}{Theorem}

\title{Exploring Information Acquisition in Social Learning: A Rational Inattention Perspective}
\name{Yiqing Lin, Zhanjiang Chen, Huisheng Wang, H.Vicky Zhao}
\address{Dept. of Automation, Tsinghua University, Beijing, China}
\begin{document}
%
\maketitle
\begin{abstract}

%

Social learning, a fundamental process through which individuals shape their beliefs and perspectives via observation and interaction with others, is critical for the development of our society and the functioning of social governance. Prior works on social learning usually assume that the initial beliefs are given and focus on the update rule. With the recent proliferation of online social networks, there is an avalanche amount of information, which may significantly influence users’ initial beliefs. In this paper, we use the rational inattention theory to model how agents acquire information to form initial beliefs and assess its influence on their adjustments in beliefs. Furthermore, we analyze the dynamic evolution of belief distribution among agents. Simulations and social experiments are conducted to validate our proposed model and analyze the impact of model parameters on belief dynamics.

\end{abstract}
\section{Introduction}
\label{sec:intro}

Social learning \cite{chamley2004rational,acemoglu2011bayesian, mossel2016efficient,  anunrojwong2018naive, bordignon2020social, bordignon2021adaptive,  chamley2013models, nedic2017fast, uribe2019non, matta2020interplay, grimm2020experiments, hu2023optimal} refers to the process through which individuals shape their beliefs and perspectives via observation and interaction with others. Understanding the fundamental mechanism behind social learning is critical for the advancement of our society and the effective functioning of social governance. 
The rise of the internet and social networks has transformed how people obtain information and communicate, and it is of paramount importance to understand how the humongous information online and the new ways of interaction affect social learning and people's beliefs. 



Many prior works on social learning study how individuals in a network communicate with their neighbors and update their beliefs. 
One of the classic belief update methods is the Bayesian-based approach \cite{chamley2004rational,acemoglu2011bayesian, mossel2016efficient,anunrojwong2018naive,bordignon2020social}, which assumes that each one is a Bayesian individual and incorporates Bayesian probability theory to describe how individuals revise their belief with new social signals. 
While the belief update rules in social learning have been widely studied, most prior works assume that each individual's initial belief is given and overlook how it is formed.
From \cite{bobkova2022two}, one of the primary ways individuals establish their initial beliefs is through the acquisition of information, such as reading the news.\footnote{Note that many factors may contribute to the formation of the initial belief, such as prior experience in related events, etc. To simplify the analysis, in this work, we only consider the impact of information acquisition and ignore all other factors. We plan to consider how other factors affect the establishment of the initial belief in our future work.} 
The work in \cite{bobkova2022two} studied information acquisition in social learning in the sequential learning scenario, where messages were transmitted from one individual to another sequentially. 
However, with the internet and online social networks (OSNs), users can access an avalanche amount of information from various sources simultaneously, and they can only absorb a small fraction due to their limited attention available \cite{sims2003implications}. Thus, the new challenge is how to model the process where humans choose information from endless sources to form their initial beliefs, and to analyze its impact on the evolution of beliefs. 

To address this challenge, we explore information acquisition in social learning from a rational inattention perspective. Rational Inattention (RI) theory \cite{sims2003implications, mackowiak2023rational, Caplin2017RationallyIB, dewan2020estimating} from behavioral economics models how individuals selectively allocate their limited attention to information, and is applied in various fields \cite{Stevensrestudrdz036, YangAcquisition19,dasgupta2018inattentive}.
It states that humans cannot process all information available but rather choose what to focus on and highlights the trade-off between getting more information for optimal outcomes and conserving attentional cost. This means that when humans acquire information, it will inevitably involve uncertainty due to limited cognitive resources.




In this work, we use the rational inattention theory to model the information acquisition stage and analyze its impact on social learning. 
First, we follow the RI theory and define individuals' selective information acquisition as a utility maximization problem that balances the benefits and costs of information acquisition. We assume that additional time incurs higher costs but yields more accurate signals.
Then, we analytically derive the dynamic evolution and statistical properties of the agents' beliefs on social networks. 
Finally, we conduct social and simulation experiments to validate our proposed model and analysis. 
 


\section{Problem Formulation}

\begin{figure}[t]
  \centering
  \includegraphics[width=1.0\linewidth]{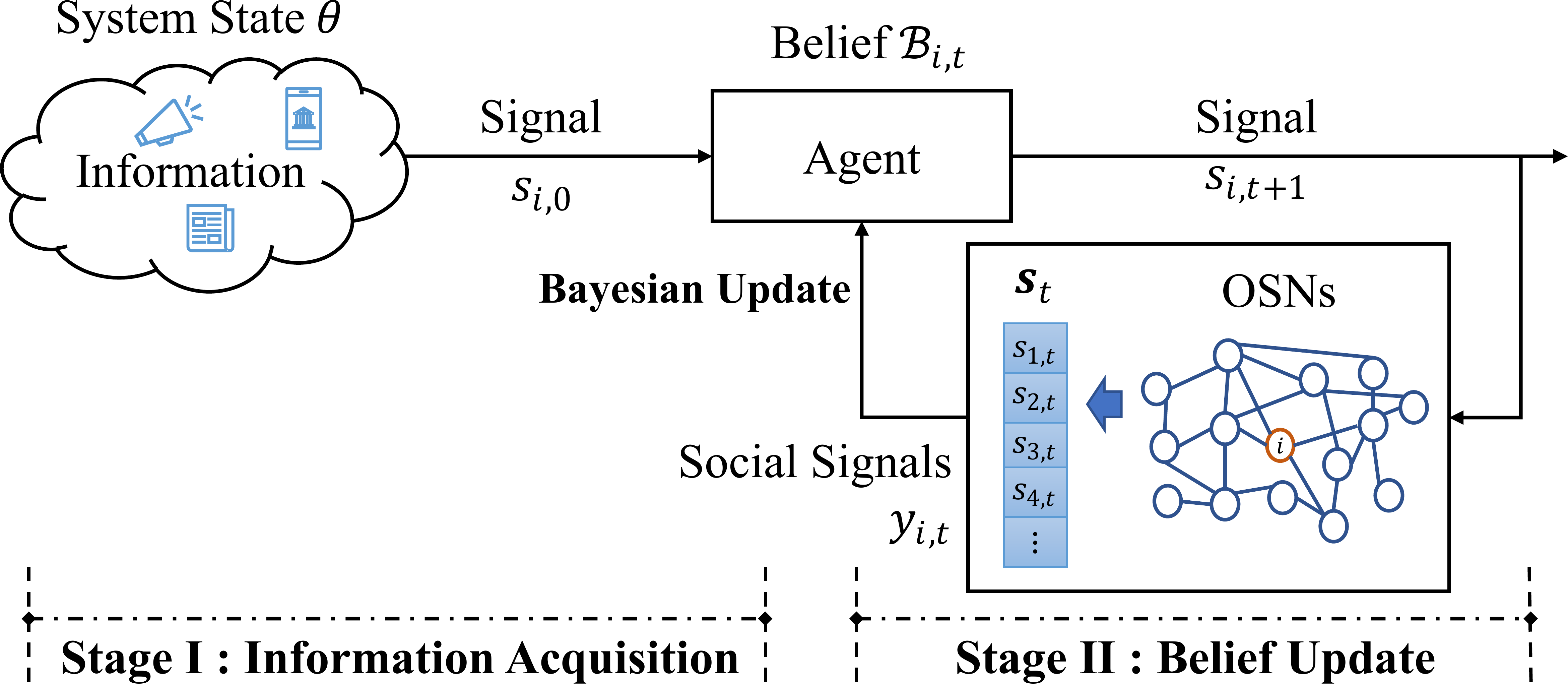}
  \caption{The proposed framework.}
  \label{fig:overview}
\end{figure}

In the following, we will use the terms ``individuals'', ``agents'' and ``users'' interchangeably. 
Assume that the network has a total of $N$ agents, and $\mathbf{W}$ is its adjacency matrix. We consider a directed graph where $\mathbf{W}$ is row-stochastic with $\sum_{j=1}^N{w_{i,j}}=1$, $w_{i,j} \geq 0$ and $w_{i,i}=0, \forall i,j$. $w_{i,j}$ quantifies agent $j$'s influence on agent $i$ in belief update. 
Assume that there is a system state $\theta$, which can be represented using a real number, and we assume it cannot be directly observed. To infer the true state $\theta$, users browse the web and/or social networks for related information 
to form their initial beliefs. Then, they exchange their opinions with neighbors and update their beliefs. There are two stages in the whole process: information acquisition and belief update, as shown in Fig. \ref{fig:overview}.


Following the RI theory, in the first stage, agent $i$ acquires information on his/her own to form his/her initial belief $\mathcal{B}_{i,0}$. As agents cannot process all information due to their limited attention, following the works in \cite{sims2003implications, mackowiak2023rational}, we model agent $i$'s initial belief $\mathcal{B}_{i,0}$ as a Gaussian distribution $\mathcal{N}(\theta, \sigma_{i,0}^2)$ to address the uncertainties in agent's belief. 
$\sigma_{i,0}^2$  quantifies agent $i$'s effort on information acquisition. A smaller $\sigma_{i,0}^2$ indicates that agent $i$ spends more time and attention in collecting and processing information at a higher cost and gets a more accurate estimate of $\theta$.
We assume that in this stage, agents collect and process information independently, and thus $\{ \mathcal{B}_{i,0} \}$ are independent of each other with different $\{ \sigma^2_{i,0} \}$. Then, agent $i$ publishes a signal $s_{i,0}$, which is a random variable following his/her belief $\mathcal{B}_{i,0}$ and shares it with his/her neighbors. 


In the second stage, at time $t+1$, given signals $\{ s_{j, t} \}$, agent $i$ first linearly combines his/her neighbors' signals and obtains the social signal $y_{i,t}=\sum_{j=1}^N w_{i,j} s_{j,t}$. Then, assuming that all agents are Bayesian individuals, given $y_{i,t}$, agent $i$ applies the Bayesian update rule to get the posterior probability of belief $\mathcal{B}_{i,t+1}$ with details in Section \ref{sec:evol}.
Similar to \cite{ma2015dynamic}, we assume that agents have a priori information of their neighbors' variances $\{ \sigma^2_{j,0} \}$, which they will use in the belief update process. 
Finally, agent $i$ exchanges the updated signal $s_{i, t+1} \sim \mathcal{B}_{i,t+1}$ with his/her neighbors and repeats the above process to infer the true state $\theta$ until it converges.


\section{Theoretical Analysis}
In this section, we theoretically analyze the formation of initial belief and its impact on social learning. 


\subsection{The Information Acquisition Stage}
\label{sec:RI}

Following the RI theory, given the true system state $\theta$, agent $i$ selects $\sigma_{i,0}^2$, that is, time/effort spent to collect and process information, to balance the cost and the benefits of information acquisition, forms the initial belief $\mathcal{B}_{i,0}$, and publishes the signal $s_{i,0} \sim {\mathcal N}(\theta, \sigma_{i,0}^2)$. \\
\textbf{Cost:} 
In our problem, following prior works in RI, we define the cost function as $C(\sigma_{i,0}^{2})$, which is a decreasing function of $\sigma_{i,0}^{2}$. That is, it takes more effort to obtain more accurate information. Our social experiments in Section \ref{sec:social-exp} suggest that the cost function can be approximated as a power function $C(x)=a_ix^{-b_i}$, where $ a_i,b_i>0$ are constants. 
\\
\textbf{Benefit:} We define the benefit as the accuracy of $s_{i,0}$, using the classic quadratic form function in RI theory $-r_i(\theta-s_{i,0})^2$. $r_i>0$ is a constant quantifying the importance of getting an accurate estimate of $\theta$ in the information acquisition stage. \\
\textbf{Utility Maximization:} Agent $i$ chooses $\sigma_{i,0}^2$ to balance the tradeoff between cost and benefit, and solves the following optimization problem:
\begin{equation}
\label{optimize}
\begin{aligned}
	&\underset{\sigma_{i,0}^{2} \in \mathcal{R}_+}{\max}\ \mathbb{E}\left[-r_i(\theta-s_{i,0})^2-a_i (\sigma_{i,0}^2)^{-b_i}\right], \\
	&\quad\mathrm{s.t.} \quad (a_i,b_i,r_i)\in\mathcal{R}_+^3.
\end{aligned}
\end{equation}
It is easy to show that the solution of (\ref{optimize}) is
\begin{equation}
\label{var}
    ({\sigma_{i,0}^{2}})^{*}={\left(\frac{r_i}{a_i b_i} \right)}^{\frac{1}{-b_i-1}}.
\end{equation}
Therefore, we get the initial belief $\mathcal{B}_{i,0} \sim \mathcal{N}(\theta,{(\frac{r_i}{a_i b_i} )}^{\frac{1}{-b_i-1}})$. 

\subsection{The Belief Update Stage}
\label{sec:evol}
Then, agents communicate with their neighbors and update their beliefs. At time $t$, agent $i$'s belief $\mathcal{B}_{i,t}$ is a Gaussian distribution $\mathcal{N}(\pi_{i,t},\sigma_{i,t}^2)$, and his/her signal published at time $t$ is $s_{i,t}$ following the same distribution. We first analyze how agent $i$ updates his/her belief, and then extend the analysis to all agents in the network. \\
\textbf{Agent $i$'s Belief Update:} 
We adopt the classical Bayesian social learning framework, assuming that each agent is a Bayesian individual. At time $t+1$, agent $i$ first averages his/her neighbors' signals and obtains the social signal $y_{i,t}$. As the analysis of the true distribution of $y_{i,t}$ is complicated, in this work, we approximate it as a Gaussian distribution $\mathcal{N}(\eta_{i,t}, {\sigma^2_{{y_i},t}})$ where 
\begin{equation}
\label{socialsignalstats}
	\eta_{i,t}=\sum_{j=1}^N w_{ij}s_{j,t},\ \mbox{and} \  {\sigma^2_{{y_i},t}}=\sum_{j=1}^N  w_{ij}^2{\sigma^2_{j,t}}.
\end{equation}
Then, agent $i$ uses the Bayesian criteria to update his/her belief and we have the following Theorem 1.
\begin{theorem}[Belief Update Rule]
The updated belief obeys $\mathcal{B}_{i,t+1} \sim \mathcal{N}(\pi_{i,{t+1}},{{\sigma^2_{i,{t+1}}}})$ where 
\begin{align}
	\pi_{i,{t+1}}&=\alpha_{i,{t}} \sum_{j=1}^N {w_{ij}s_{j,t}}+\left(1-\alpha_{i,t} \right)\pi_{i,t}\ , \cr
	{\sigma^{2}_{i,{t+1}}}&=\frac{{\sigma^2_{i,{t}}}{\sigma^2_{{y_i},t}}}{{\sigma^2_{i,t}}+{\sigma^2_{{y_i},t}}}=\frac{\sum_{j=1}^N  w_{ij}^2{\sigma^2_{i,{t}}}{\sigma^2_{{j},t}}}{{\sigma^2_{i,t}}+\sum_{j=1}^N  w_{ij}^2{\sigma^2_{j,t}}}, \cr
\text{and} \ \alpha_{i,t}&=\frac{{\sigma^2_{i,{t}}}}{{\sigma^2_{i,t}}+{\sigma^2_{{y_i},t}}}=\frac{{\sigma^2_{i,{t}}}}{{\sigma^2_{i,t}}+\sum_{j=1}^N  w_{ij}^2{\sigma^2_{j,t}}}.
\label{evolu_single}
\end{align}
\end{theorem}
From (\ref{evolu_single}), the updated mean $\pi_{i,{t+1}}$ is a weighted average of the mean $\pi_{i,{t}}$ at time $t$ and the mean of the social signal $\eta_{i,t}$, and the weight $\alpha_{i,t}$ is determined by the variance $\sigma_{i,t}^2$ at time $t$ and the social signal's variance $\sigma^2_{{y_i},t}$. When $\sigma^2_{{y_i},t}$ is large, that is, his/her neighbors' signals are noisy, agent $i$ relies more on his/her own belief with a smaller $\alpha_{i,t}$. Otherwise, his/her neighbors' estimates of $\theta$ are more reliable, and agent $i$ puts a larger weight on the social signal $y_{i,t}$ with a larger $\alpha_{i,t}$.


Note that in (\ref{evolu_single}), $s_{j,t}$ is a random variable, and thus, the mean $\pi_{i,{t}}$ is also a random variable. Therefore, when analyzing the statistical properties of $\mathcal{B}_{i,t+1}$ and $s_{i,t+1}$, we also need to consider the mean/variance of $\pi_{i,{t}}$. Let $\delta_{i,t}$ be the variance of $\pi_{i,{t}}$. Overall, 
the mean and the variance of $s_{i,t+1}$ are
\begin{align}
\label{vars}
    \mathbb{E}(s_{i,t+1})&=\mathbb{E}(\pi_{i,t+1})=\theta \; \mbox{and}\\
    \mathbb{D}(s_{i,t+1})&=\left(1-\alpha_{i,t}\right)^2\delta^2_{i,{t}}+\left(\alpha_{i,t}\right)^2\sum_{j=1}^N w_{ij} (\delta^2_{j,{t}}+\sigma_{j,t}^2) \cr
		 & +2\alpha_{i,t}(1-\alpha_{i,t}) Cov(\sum_{j=1}^N w_{ij} s_{j,t},\pi_{i,t}) +\sigma_{i,{t+1}}^2, \nonumber
\end{align}
respectively, 
where the covariance can be calculated recursively.\\

\noindent
\textbf{Belief Update of the Whole Network:} We extend the above analysis to a network with $N$ agents. $\bm{s}_t=[ s_{1,t}, s_{2,t}, \cdots, s_{N,t}]^\top$ is the vector of all agents' signals at time $t$, and the mean vector of all agents' beliefs is
$\boldsymbol{\pi}_t=[\pi_{1,t} , \pi_{2,t} , \cdots , \pi_{N,t}]^\top$. For simplicity, we define $\boldsymbol{\sigma}^{-2}_t=[
        \sigma^{-2}_{1,t} , \sigma^{-2}_{2,t}, \cdots , \sigma^{-2}_{N,t}
    ]^\top$, and the variance of social belief as $\boldsymbol{\sigma}_{s,t}^{-2} =\mathbf{ W}^{\circ -2}  \boldsymbol{\sigma}_{t}^{-2}$, where for $\mathbf{ W}^{\circ -2}$, its element at row $i$ and column $j$ is ${w_{ij}^{-2}}$. The weight coefficient matrix is $\mathbf{A}_t=\text{diag}([ \alpha_{1,t} , \alpha_{2,t} , \cdots , \alpha_{N,t} ]) $. Then for all agents in the network, \textbf{Theorem 1} can be extended to the following vector-matrix form
\begin{align}
\label{model}
		\boldsymbol{\pi}_{t+1} &=\mathbf{A}_t \mathbf{W} \bm{s}_t + \left(\mathbf{I}-\mathbf{A}_t \right)\boldsymbol{\pi}_{t}, \\
		\boldsymbol{\sigma}^{-2}_{t+1}& =\sum_{\tau=1}^{t} \boldsymbol{\sigma}_{s,\tau}^{-2} + \boldsymbol{\sigma}_{0}^{-2}, \nonumber
\end{align}
where $\boldsymbol{\sigma}_{0}^{-2} =\left[ (\sigma_{1,0}^{-2})^*, \cdots, (\sigma_{N,0}^{-2})^* \right]^\top$. 

In (\ref{model}), $\bm{s}_t$ and $\boldsymbol{\pi}_t$ are random vectors, and we can have the following Theorem 2 and 3.
\begin{theorem}
$\boldsymbol{\pi}_{t}$ is a mean stationary process with $ \mathbb{E}(\bm{s}_t)=\mathbb{E}(\boldsymbol{\pi}_{t})=\theta \cdot \mathbf{1}$.
\end{theorem}

It states that the mean of an agent's belief at all times remains the same and equals the system state $\theta$. Also, despite being a mean stationary process, $\boldsymbol{\pi}_t$ is not a martingale unless $\mathbf{W}=\mathbf{I}$. In reality, such networks do not exist, and therefore, agents will be influenced by their neighbors, not satisfying the martingale property.

\begin{theorem}
Given the signal vector $\bm{s}_{t}$, we have 
\begin{align}
\label{signalvarvec}
		  &\mathbb{D}(\bm{s}_{t})= \left[ \mathbb{D}(s_{1,t}), \cdots, \mathbb{D}(s_{N,t}) \right]^\top \\ &= 
    \left(\mathbf{I}-\mathbf{A}_{t-1}\right)^2\boldsymbol{\delta}^2_{t-1}+\mathbf{A}_{t-1}^2\mathbf{W}^{\circ2}\left(\boldsymbol{\delta}^2_{t-1}+\boldsymbol{\sigma}^2_{t-1}\right)\cr
		& \;\;\; +2\mathbf{A}_{t-1}\left(\mathbf{I}-\mathbf{A}_{t-1}\right)\mathbf{W}\mathbf{Cov}\left(\boldsymbol{\pi}_{t-1},\boldsymbol{\pi}_{t-1}\right) +\boldsymbol{\sigma}^2_{t}. \nonumber
\end{align}
\end{theorem}
From (\ref{model}), $\boldsymbol{\sigma}^2_{t}$ decreases as $t$ increases and agents learn from their neighbors. However, as $\bm{s}_{t}$ and $\boldsymbol{\pi}_{t}$ are random vectors, from (\ref{signalvarvec}), $\mathbb{D}(\bm{s}_{t})$ also depends on $\boldsymbol{\sigma}^2_{0}$ and does not necessarily always decrease with time.

\section{Experiments}

\subsection{Social Experiments}
\label{sec:social-exp}


First, we design a social experiment to validate our analysis of the information acquisition stage in Section \ref{sec:RI}. 
137 subjects are recruited to play the cognitive game in \cite{das2013debiasing}, where each subject is shown images with red and blue balls and is asked to estimate the proportion of red ones.
The experiment involves multiple rounds with different configurations, including different numbers of balls and different proportions of red balls. The red balls' positions are randomized to prevent pattern learning. We conduct two experiments to verify our cost function and our analysis of the initial belief's variance in (\ref{var}). \\
\textbf{Experiment 1: }
In this experiment, each image is displayed for $T \in \{10,20,40,60,90\}$ seconds. Without informing the subjects, the configuration of the images is the same under different time limits and only the positions of red balls are shuffled. For each image, each subject reports his/her estimate of the percentage of red balls. 
\begin{figure}[htb]
  \begin{minipage}[b]{.49\linewidth}
    \centering
    \hspace{-0.5cm}
    \centerline{\includegraphics[width=1.025\linewidth]{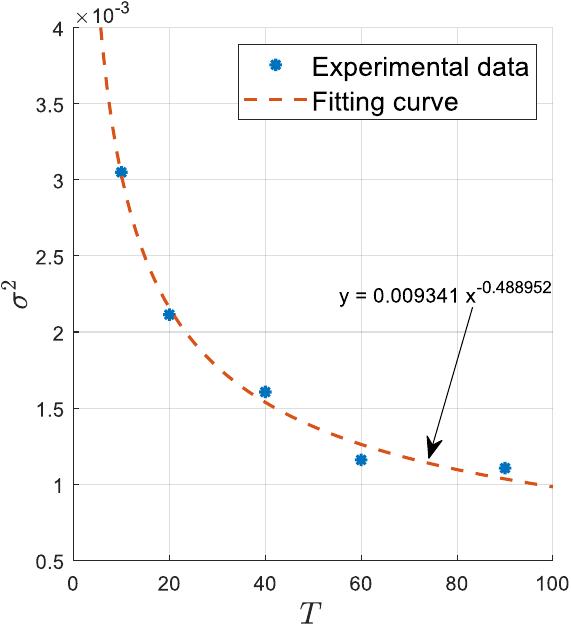}}
    \centerline{(a)}\medskip
  \end{minipage}
  \begin{minipage}[b]{0.49\linewidth}
    \centering
    \centerline{\includegraphics[width=1.1\linewidth]{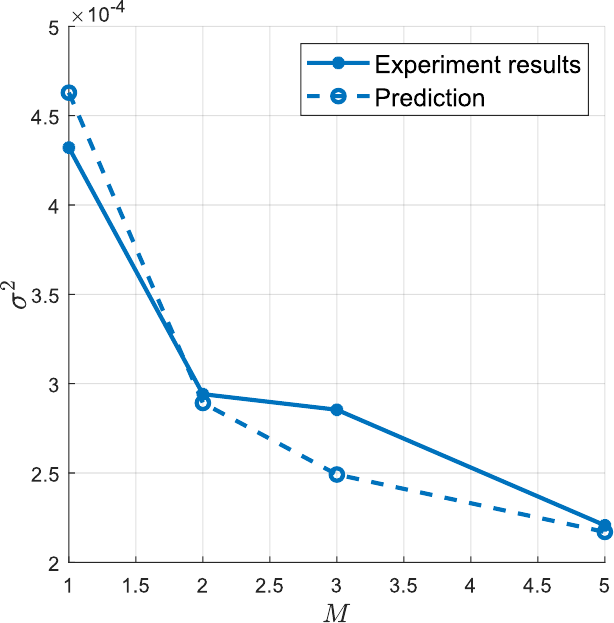}}
    \centerline{(b)}\medskip
  \end{minipage}
  \caption{Social experiments results. (a) Fitting results of the cost function, and (b) $\sigma^2$ with different
  reward amounts.}
  \label{fig:social}
\end{figure}
As the social experiment data (the number of subjects) is relatively limited, we assume that the parameters $a$ and $b$ are the same for all subjects. 
We define $T$ as the cost, and for each $T$, we calculate the variance of the reports from all subjects $\sigma^2$. Then, given all pairs of $(T, C(\sigma^2))$, we use regression to find the optimal parameters $a$ and $b$. From the fitting results shown in Fig \ref{fig:social}a, our estimated cost function fits the experimental data very well, and when given more time (a higher cost), the subjects' estimates are more accurate, which aligns with our hypothesis. \\
\textbf{Experiment 2:}
In this experiment, we verify our analysis of the information acquisition stage, specifically, the variance of the initial belief (\ref{var}). In this experiment, we do not limit the display time of the images, and the subjects 
choose how long they wish to view the image before submitting their estimates. To verify (\ref{var}), we vary the reward amount $M\in\{1,2,3,5\}$ the subjects receive once they finish the experiment. Here we assume that for different reward amounts, the parameter $r$ in (\ref{var}) are different; while for the same $M$, all subjects share the same $r$. To learn $r$ for different $M$, we divide all subjects into two groups: the training group and the testing group. Given the estimated parameters $a$ and $b$ from the first experiment, we use data from the training group to estimate $r$ for different $M$, assuming that $\sigma^2$ and $r$ satisfy (\ref{var}). Then, given the estimated $r$, we use (\ref{var}) to predict $\sigma^2$ for the reported signals from the testing group, and compare it with the ground truth value. The results are shown in Fig. \ref{fig:social}b. From there, we can see that our analysis in (\ref{var}) can accurately predict $\sigma^2$ with different reward amounts, and the average relative prediction error is 6.63\%, verifying our analysis in Section \ref{sec:RI}.

\subsection{Simulation Experiments}
\label{sec:simu-exp}

\begin{figure}[htb]
  \begin{minipage}[b]{.48\linewidth}
    \centering
    \hspace{-0.2cm}
    \centerline{\includegraphics[width=1.1\linewidth]{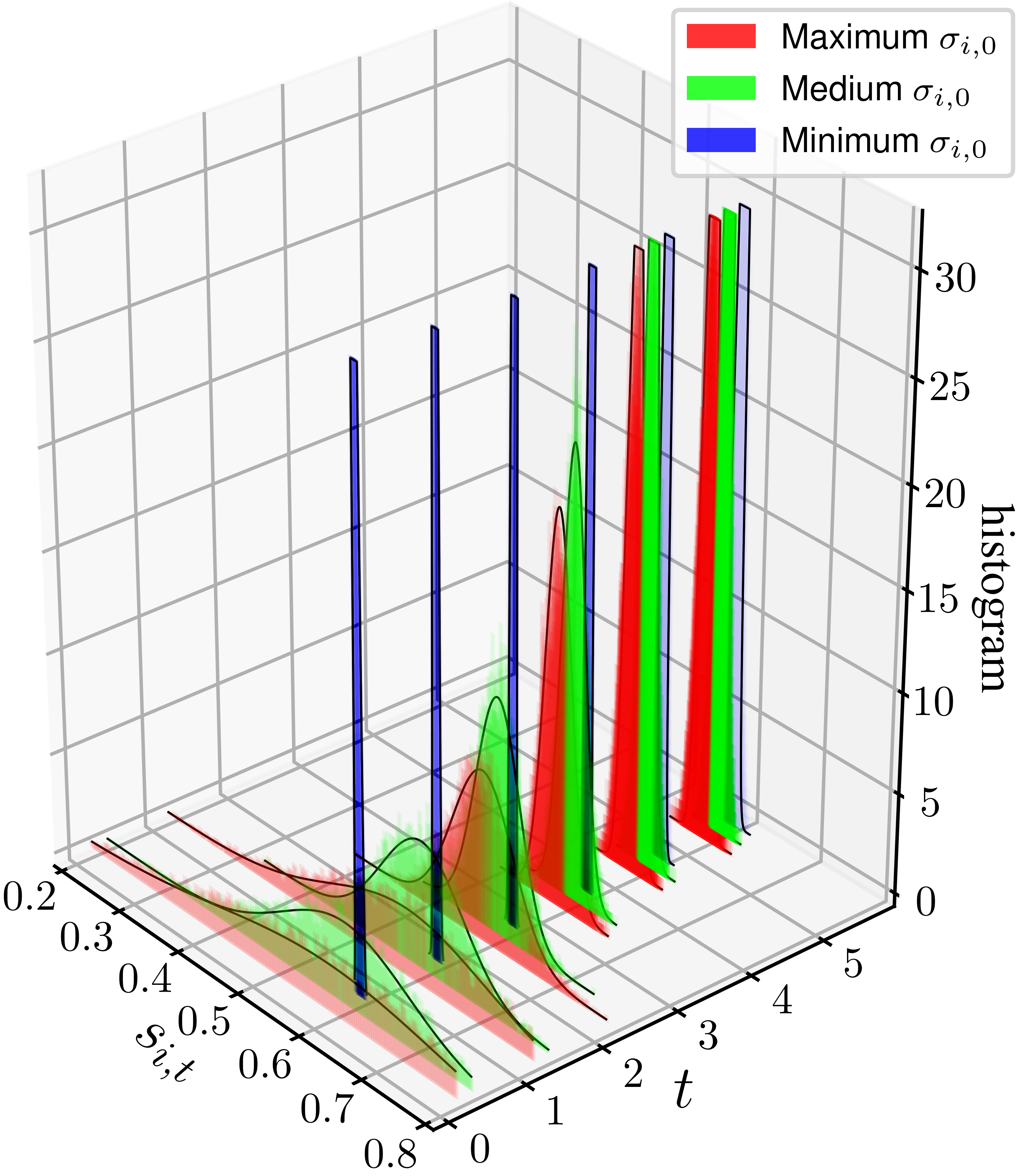}}
    \centerline{(a)}\medskip
  \end{minipage}
  \begin{minipage}[b]{0.50\linewidth}
    \centering
    \centerline{\includegraphics[width=0.95\linewidth]{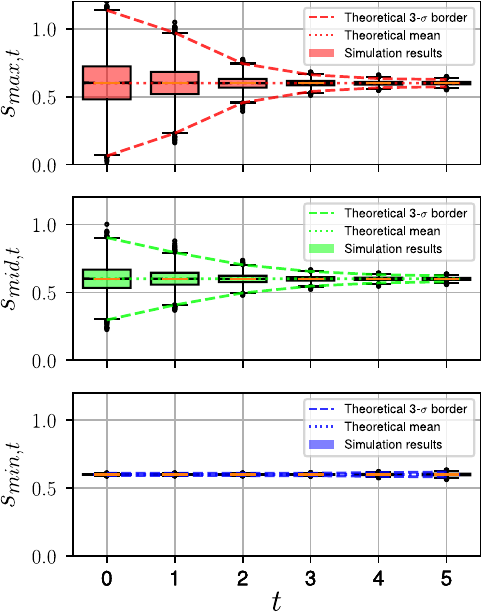}}
    \centerline{(b)}\medskip
  \end{minipage}
   \caption{Theoretical and simulation results for 3 agents with different initial beliefs. 
(a) Histograms of $s_{i,t}$ of the 3 agents, and
(b) boxplots of $s_{i,t}$ of the three agents.} 
  \label{fig:sim-expsvg}
\end{figure}

We use simulations to verify our theoretical analysis in Section \ref{sec:evol}. All experiments are performed on a randomly generated Barabási-Albert scale-free network \cite{barabasi1999emergence} with $N=100$ agents and an average degree of $k=3$. 
We set $\theta=0.6$ as an example and randomly generate the initial variances ${\sigma_{i,0}^2}$ from the interval $[0.009, 0.18]$ for all agents.
We run the simulation 10,000 times and show the results in Fig. \ref{fig:sim-expsvg}. For better visualization, we choose three agents with the maximum, median, and minimum $\sigma_{i,0}^2$, and plot the evolution of their signals $s_{max,t},s_{mid,t},s_{min,t}$ in red, green, and blue, respectively. 

Fig. \ref{fig:sim-expsvg}a plots of the histograms of $s_{max,t},s_{mid,t},s_{min,t}$ over time. 
From Fig. \ref{fig:sim-expsvg}a, all three agents' beliefs gradually converge to the true system state $\theta=0.6$ as time goes, verifying \textbf{Theorem 2}. Comparing the three subfigures in Fig. \ref{fig:sim-expsvg}b, we can see that, for agents with smaller $\sigma_{i,0}^2$, their signals' variance is smaller as they are less influenced by others during the belief update process. 
The dash lines in Fig. \ref{fig:sim-expsvg}b represent the 3$\sigma$ borderlines ($\theta \pm 3\sqrt{\mathbb{D}(s_{i,t})}$) obtained from the theoretical analysis of the variance in \textbf{Theorem 3}, which is consistent with the 3$\sigma$ upper and lower bounds obtained from the simulation results. This demonstrates the correctness of our theoretical analysis.
In addition, after enough rounds of updates, all agents' beliefs have a small variance, showing that social learning can help infer the true state value, even for those whose initial beliefs have a large variance. 

In summary, when exposed to an infinite amount of information, individuals form their initial beliefs with uncertainty due to the limited attention available.
Our theoretical analysis and experimental results show that the initial beliefs' variance affects social learning, in particular, the weights they use in belief updates.

\section{Conclusion}

In this work, we study how the avalanche amount of information online affects agents' initial beliefs and social learning. We use the rational inattention theory to model agents' selective information acquisition as a utility maximization problem that balances the trade-off between getting a more accurate estimate of the system state and spending more time and effort to collect and process information. We then theoretically analyze the dynamics of belief distribution, and our study shows that the initial beliefs will affect the weights agents use in belief updates. If agents put in more effort to collect/process information, they will be less influenced by others in belief updates. Otherwise, they will rely more on their neighbors to learn the true system state. Social experiments and simulation results validate our model and theoretical analysis. 



\vfill\pagebreak
\bibliographystyle{IEEEbib}
\bibliography{strings,refs}

\end{document}